\documentclass[twocolumn,showpacs,preprintnumbers,amsmath,amssymb,prl,superscriptaddress]{revtex4-1}
\usepackage{mathrsfs}
\usepackage{graphicx}
\usepackage{dcolumn}
\usepackage{bm}
\usepackage{amsmath}
\usepackage{amsfonts}
\usepackage{color}

\begin{document}

\title{Quantized topological magnetoelectric effect of the zero-plateau quantum anomalous Hall state}
\author{Jing Wang}
\affiliation{Department of Physics, McCullough Building, Stanford University, Stanford, California 94305-4045, USA}
\affiliation{Stanford Institute for Materials and Energy Sciences, SLAC National Accelerator Laboratory, Menlo Park, California 94025, USA}
\author{Biao Lian}
\affiliation{Department of Physics, McCullough Building, Stanford University, Stanford, California 94305-4045, USA}
\author{Xiao-Liang Qi}
\affiliation{Department of Physics, McCullough Building, Stanford University, Stanford, California 94305-4045, USA}
\author{Shou-Cheng Zhang}
\affiliation{Department of Physics, McCullough Building, Stanford University, Stanford, California 94305-4045, USA}
\affiliation{Stanford Institute for Materials and Energy Sciences, SLAC National Accelerator Laboratory, Menlo Park, California 94025, USA}

\begin{abstract}
Topological magnetoelectric effect in a three-dimensional topological insulator is a novel phenomenon, where an electric field induces a magnetic field in the same direction, with a universal coefficient of proportionality quantized in units of $e^2/2h$. Here we propose that the topological magnetoelectric effect can be realized in the zero-plateau quantum anomalous Hall state of magnetic topological insulators or ferromagnet-topological insulator heterostructure. The finite-size effect is also studied numerically, where the magnetoelectric coefficient is shown to converge to a quantized value when the thickness of topological insulator film increases. We further propose a device setup to eliminate the non-topological contributions from the side surface.
\end{abstract}

\date{\today}

\pacs{
        73.43.-f  
        73.20.-r  
        85.75.-d  
      }

\maketitle

The search for topological quantum phenomena has become an important goal in condensed matter physics. Topological phenomena in the physical systems are determined by topological structures and are thus universal and robust against perturbations, and the electromagnetic response is usually exactly quantized~\cite{thouless1998}. Two well-known examples of topological quantum phenomena are the flux quantization in superconductors~\cite{byers1961} and Hall conductance quantization in the quantum Hall effect (QHE)~\cite{thouless1982}. The remarkable observation of such topological phenomena is that the quantization is exact, which provide the precise values of fundamental physics constants such as Plank's constant $h$~\cite{mohr2008}.

The recent discovery of the time-reversal ($\mathcal{T}$) invariant (TRI) topological insulator (TI) brings the opportunity to study a large family of new topological phenomena~\cite{hasan2010,qi2011}. The electromagnetic
response of a three-dimensional (3D) insulator is described by the topological $\theta$ term~\cite{qi2008,essin2009,unit} of the form
\begin{equation}
\mathcal{S}_{\theta} = \frac{\theta}{2\pi}\frac{e^2}{h}\int d^3xdt\mathbf{E}\cdot\mathbf{B},
\end{equation}
together with the ordinary Maxwell terms. Here $\mathbf{E}$ and $\mathbf{B}$ are the conventional electromagnetic fields inside the insulator, $e$ is the charge of an electron, and $\theta$ is the dimensionless pseudoscalar parameter describing the insulator, which refers to the axion field in particle physics~\cite{wilczek1987}. Under the periodic boundary condition, all physical quantities are invariant if $\theta$ is shifted by integer multiples of $2\pi$. Therefore, all TRI
insulators are described by either $\theta=0$ or $\theta=\pi$ (modulo $2\pi$). TIs are defined by $\theta=\pi$, which cannot be connected continuously to trivial insulators, defined by $\theta=0$, by TRI perturbations. With open boundary condition, the effective action is reduced to a (2+1)D Chern-Simons term on the surface, which describes a surface QHE with half-quantized surface Hall conductance~\cite{qi2008}.
Such a topological $\theta$ term with a universal value of $\theta=\pi$ in TIs leads to a magnetoelectric effect with coefficient quantized in units of $e^2/2h$, known as the topological magnetoelectric effect (TME), i.e., an electric field can induce a magnetic polarization, whereas a magnetic field can induce an electric polarization. To obtain the quantized TME in TIs, as is first suggested in Ref.~\cite{qi2008}, one must fulfil the following stringent requirements. First, introduce a $\mathcal{T}$-breaking surface gap by ferromagnetic (FM) ordering, where the magnetization of FM points inward or outward from the surface. Second, finely tune the Fermi level into the magnetically induced surface gap and keep the bulk truly insulating. Third, the film of TI material should be thick enough to eliminate the finite-size effect, therefore the TME is exactly quantized. Several other theoretical proposals~\cite{qi2009b,maciejko2010,tse2010,nomura2011} have been made to realize the TME; however, observing the TME in TIs experimentally is still challenging.

\begin{figure}[t]
\begin{center}
\includegraphics[width=3.3in]{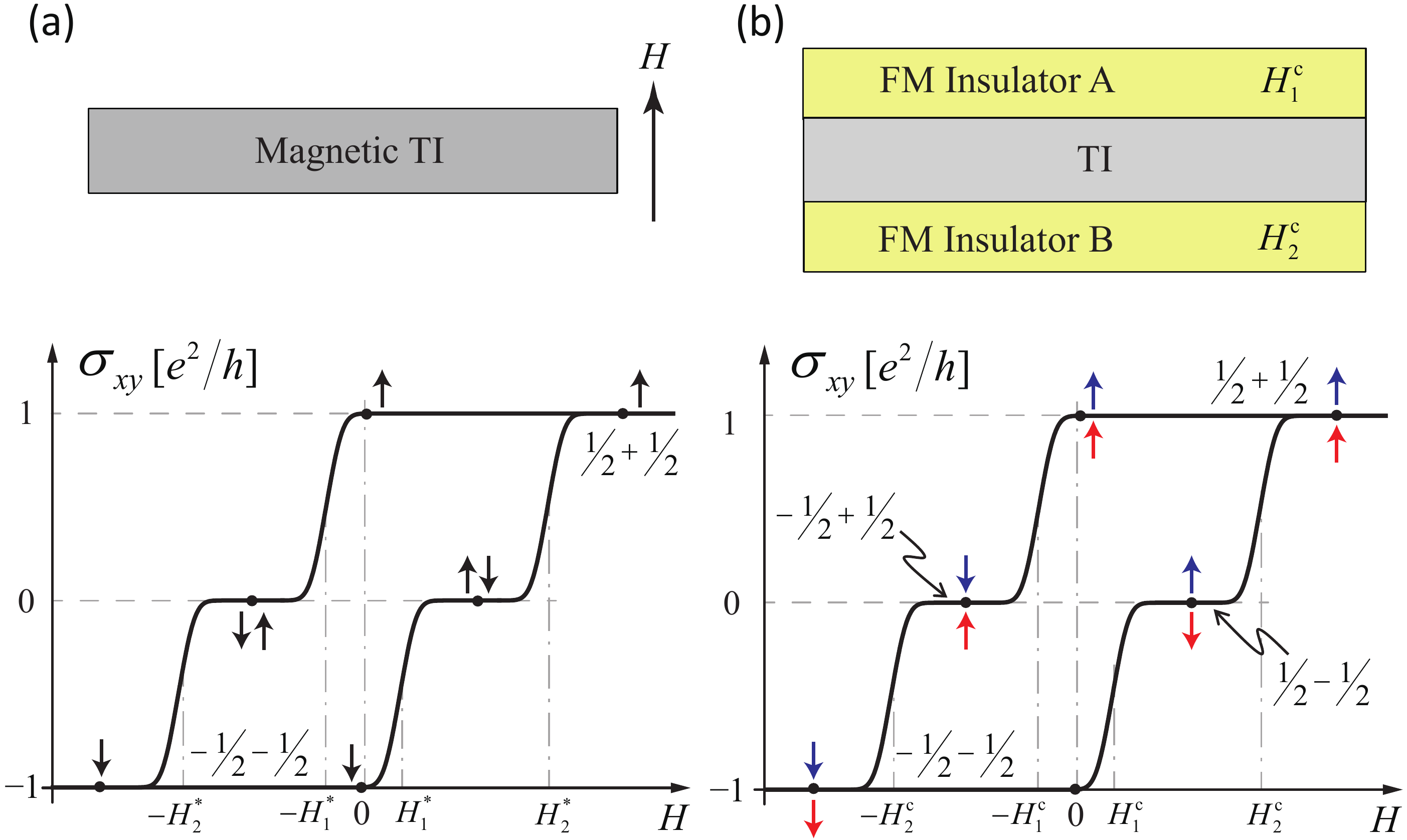}
\end{center}
\caption{(color online). Zero-plateau QAH state and magnetic field dependence of $\sigma_{xy}$. (a) Magnetic TI in an external field $H$, and sketch of $\sigma_{xy}$ as a function of $H$. $\sigma_{xy}=0$ plateau occurs at the coercivity. The arrow indicates the magnetization direction. (b) FM-TI-FM heterostructure, and $\sigma_{xy}$ vs $H$. The zero-plateau appears in the hysteresis loop due to different $H^c_1$ and $H^c_2$ of FM A and B. The blue and red arrows indicate the magnetization direction of FM A and B, respectively.}
\label{fig1}
\end{figure}

In this paper, we propose to realize TME effect in the newly discovered quantum anomalous Hall (QAH) state~\cite{chang2013b,wang2014b}. Recently, a new zero-plateau QAH state in a magnetic TI has been theoretically predicted~\cite{wang2014a} and experimentally realized~\cite{fengy2015,kou2015}. The magnetic TI studied in the QAH experiment develops robust FM at low temperature. In the magnetized states, the magnetic domains of the material
are aligned to the same direction, and the system is in a QAH state with a single chiral edge state propagating along the sample boundary, where the Hall conductance $\sigma_{xy}$ is quantized to be $\pm e^2/h$. The zero-plateau state, on the contrary, appears around the coercivity when the magnetic domains reverse, where $\sigma_{xy}$ shows a well-defined zero-plateau over a range of magnetic field around coercivity while longitudinal conductance $\sigma_{xx}\rightarrow0$ [shown in Fig.~\ref{fig1}a]. In such a state, the Fermi level is in the magnetization induced surface gap, fulfilling the first two conditions above, providing a good platform to observe the TME effect as we will discuss in details below. However, due to finite thickness in magnetic TI, the TME is non-quantized. Therefore, we further propose to realize the \emph{quantized} TME effect in the zero-plateau QAH state of the FM-TI-FM heterostructure as shown in Fig.~\ref{fig1}b, where an in-plane ac magnetic field induces an electric current in the same direction, or an in-plane electric field induces a magnetic field. The finite-size effect is also studied numerically, where the TME coefficient is shown to converge to a quantized value when the thickness of TI film increases. Finally, we propose a device setup where the non-topological contribution from the side surface is negligible.

\begin{figure}[t]
\begin{center}
\includegraphics[width=3.3in]{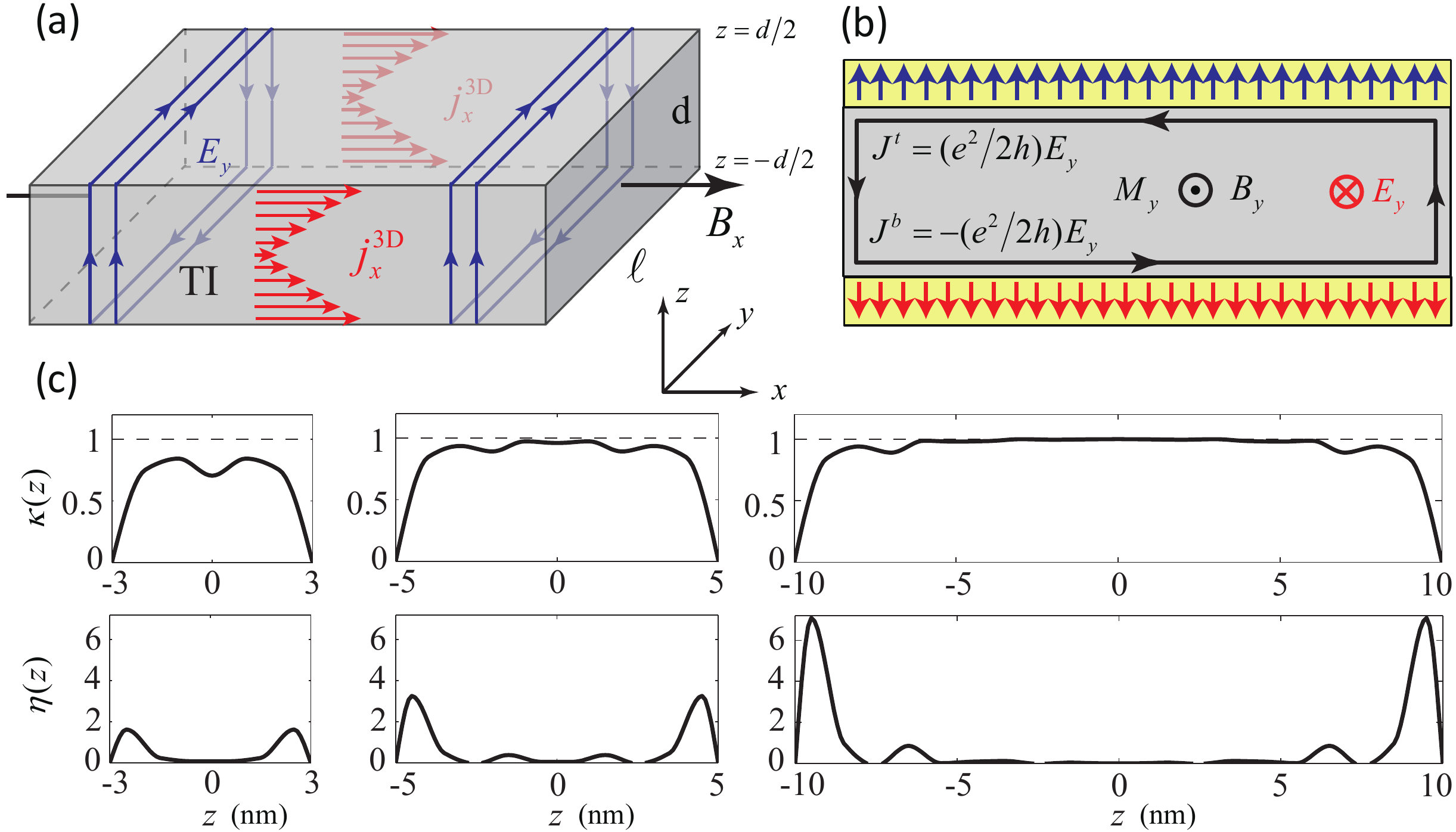}
\end{center}
\caption{(color online). TME effect. (a) Illustration of the ac electric current induced by an ac magnetic field $B_x$, the current density $j_x^{\text{3D}}(z)$ is defined in Eq.~(\ref{3Dcurrent}). (b) Magnetic field $B_y$ induced by applying an electric field $E_y$ through a capacitor. $E_y$ (with direction into the paper) will induce Hall currents $\mathbf{J}^t$ and $\mathbf{J}^b$ for antiparallel magnetization. (c) The functions $\kappa(z)$ and $\eta(z)$ for different thickness $6$, $10$, $20$~QL. Each QL is about $1$~nm thick.}
\label{fig2}
\end{figure}

The TME described by the topological $\theta$ term implies that a quantized magnetic polarization is induced by an electric field, given by
\begin{equation}
\mathbf{M}=-\frac{\theta}{2\pi}\frac{e^2}{h}\mathbf{E}.
\end{equation}
Such response can be understood in terms of a surface Dirac fermion picture. With antiparallel magnetization of the two FM layers as shown Fig.~\ref{fig2}b, an in-plane electric field $\mathbf{E}=E_y\hat{\mathbf{y}}$ induces the Hall currents $\mathbf{J}^t=\sigma_{xy}^t\hat{\mathbf{z}}\times\mathbf{E}$ and $\mathbf{J}^b=\sigma_{xy}^b\hat{\mathbf{z}}\times\mathbf{E}$ on top ($z=d/2$ and denoted as superscript $t$) and bottom ($b$ and $z=-d/2$) surfaces, respectively. Since the surface massive Dirac fermion gives rise to half-integer Hall conductance $\sigma_{xy}^t=-\sigma^b_{xy}=(\theta/2\pi)(e^2/h)$, the currents $\mathbf{J}^t=-\mathbf{J}^b$ are opposite and form a circulating total current.
Inside the sample, such a circulating current can be viewed equivalently as the surface bound current generated by a constant magnetization
$\mathbf{M}=-J^t\hat{\mathbf{y}}=-(\theta/2\pi)(e^2/h)\mathbf{E}$. Therefore, the TME essentially originates from the half-quantized surface Hall conductance.

First we examine theoretically the TME in the zero-plateau QAH state observed in experiments. The low-energy physics of this system consists of Dirac-type surface states only~\cite{wang2014a,wang2015}. At the coercivity, both \emph{random} magnetic domains that formed in the sample, and the exchange field $\Delta$ introduced by the FM ordering are spatially inhomogeneous. The 3D spatial average of exchange field vanishes ($\langle\Delta\rangle_{\text{ave}}=0$).
However, due to an unavoidable top-bottom asymmetry, the top and bottom surfaces may feel an opposite nonzero exchange field $\Delta^t=-\Delta^b=\Delta_0$. In this case, the zero-plateau state is described by the mean field effective model which has the generic form as $\mathcal{H}_0=k_y\sigma_1\otimes\tau_3-k_x\sigma_2\otimes\tau_3+\Delta_0\sigma_3\otimes\tau_3+m1\otimes\tau_1$.
with the basis of $|t\uparrow\rangle$, $|t\downarrow\rangle$, $|b\uparrow\rangle$ and $|b\downarrow\rangle$, where $\uparrow/\downarrow$ represent
the spin up/down states, respectively. $\sigma_i$ and $\tau_i$ ($i=1,2,3$)
are Pauli matrices acting on spin and layer, respectively. $m$ describes the hybridization
between the top and bottom surface states. If $m=0$, due to the opposite half-integer Hall conductance contributions from the top and bottom surfaces $\sigma_{xy}^t=-\sigma_{xy}^b=\text{sgn}(\Delta_0)(e^2/2h)$, the system has $\sigma^{\text{tot}}_{xy}=0$, which gives rise to a quantized TME as discussed above. However, a nonzero $m$ will mix the circulating current $\mathbf{J}^t$ and $\mathbf{J}^b$, therefore, the TME is no longer quantized. In reality, the exchange field depends very much on the microscopic details of the randomness in magnetic domains. However, we emphasize that the TME of the zero-plateau state in a magnetic TI is in general nonzero and non-quantized.

To realize the quantized TME effect, the TI film should be thick enough so that the hybridization between top and bottom surfaces is negligible. Therefore, we propose that a quantized TME can be realized in the zero-plateau state of FM-TI-FM structure as shown in Fig.~\ref{fig1}b. The FM insulators A and B have different coercivity $H^c_1$ and $H^c_2$, respectively. Assume that both FM A and B have an out-of-plane magnetic easy axis, and the same sign of the exchange coupling parameter to TI surface states. When A and B have antiparallel magnetization, the system is in a zero-plateau QAH state with $\sigma^{\text{tot}}_{xy}=0$, which is contributed by $\sigma_{xy}^t+\sigma_{xy}^b$ as $(1/2-1/2)(e^2/h)$ or $(-1/2+1/2)(e^2/h)$. Such a magnetization configuration can be easily achieved in the hysteresis loop by an external field $H$ with $H^c_1<|H|<H^c_2$, and then remove $H$. Experimentally, to achieve the TME in this setup, a good proximity between FM and TI is necessary. TI material can be chosen as Bi$_y$Sb$_{1-y}$Te$_3$, where the Dirac cone of the surface states
is observed to be located in the bulk band gap~\cite{zhang2011}. The candidate FM materials are Cr$_2$Ge$_2$Te$_6$ (CGT), Cr$_x$(Bi,Sb)$_{2-x}$Te$_3$ (CBST) with $0.3<x<0.46$ and V$_x$(Bi,Sb)$_{2-x}$Te$_3$ (VBST). All of them are FM insulators with an out-of-plane easy axis, and have good lattice match with Bi$_2$Te$_3$ family materials. CGT is a soft FM insulator with $T_c\sim61$~K and $H_c<100$~Oe~\cite{ji2013}, and it also shows good proximity with Bi$_2$Te$_3$~\cite{alegria2014}. CBST with $0.3<x<0.46$ is a FM insulator with $T_c=40$-$90$~K and $H_c\sim1.0\times10^3$~Oe~\cite{ke_note}. VBST with $0.1<x<0.3$ is a FM insulator with $T_c=30$-$100$~K and $H_c\sim1.0\times10^4$~Oe~\cite{chang2015}.

\paragraph{TME}

As we discussed previously, an electric field will induce a topological contribution to bulk magnetization. From the constituent equation $\mathbf{H}=\mathbf{B}/\mu-\mathbf{M}$, with $\mathbf{H}=0$ and $\mathbf{B}$ continuous, we have on the middle of side surface (parallel to $\hat{\mathbf{z}}$) $\mathbf{B}=-\mu(e^2/2h)E_y\hat{\mathbf{y}}$. Here $\mu$ is the material-dependent magnetic permeability. Taking $\mu\approx\mu_0$, $E_y=10^5$~V/m, we get the magnitude of magnetic field $2.43\times10^{-6}$~T, which is easily detectable by present superconducting quantum interference devices (SQUID). The stray magnetic field effect can be well separated from the quantized TME by ac modulation of the electric field and phase-locking detection, where the ac frequency is quasi-static around 10-100~Hz. Moreover, a gradiometer sensor in SQUID could also screen the homogeneous stray field.

\begin{figure}[t]
\begin{center}
\includegraphics[width=3.35in]{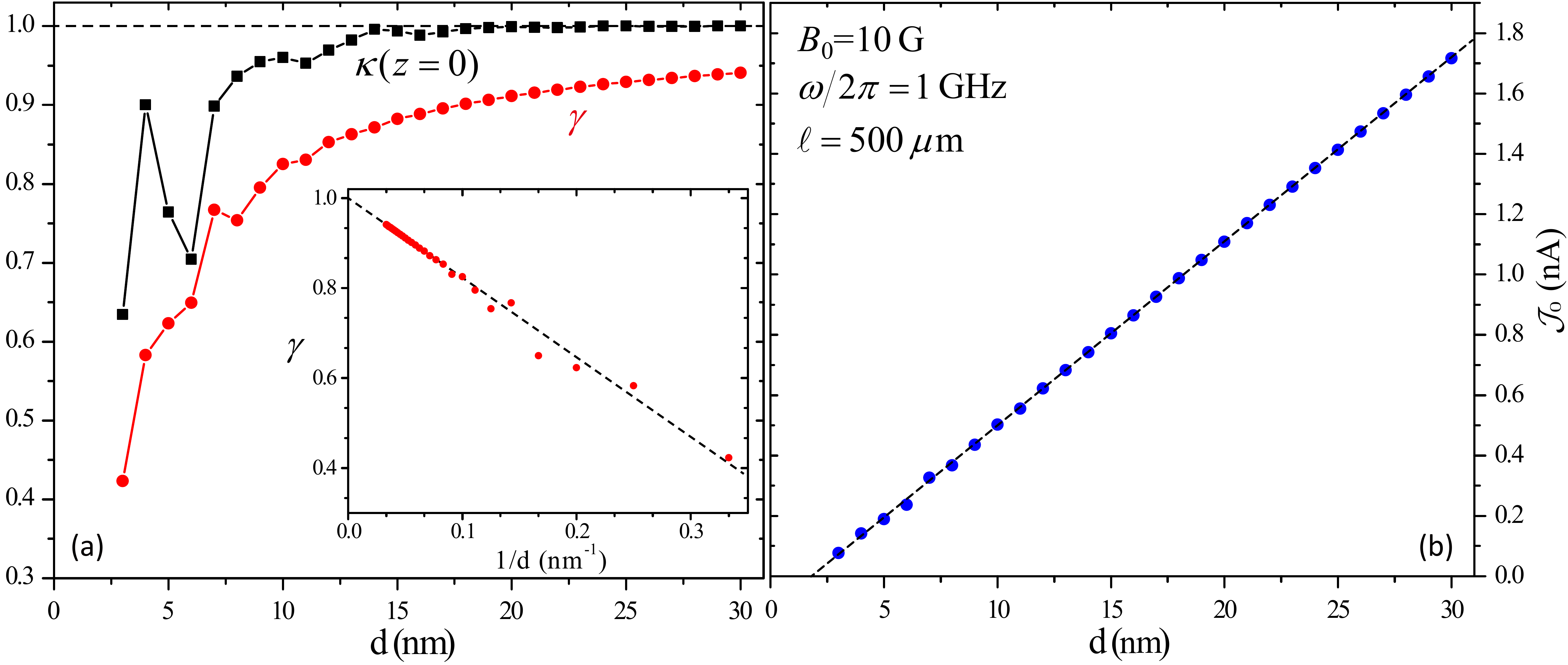}
\end{center}
\caption{(color online). Finite-size effect of TME. (a) The $\gamma$ and $\kappa(z=0)$ as a function of $d$. The inset shows $\gamma$ plotted vs the inverse of thickness $1/d$. (b) The current amplitude $\mathcal{J}_0$ scales linearly with $d$. Here $\theta/\pi=\gamma$.}
\label{fig3}
\end{figure}

The TME also indicates the induction of a parallel polarization current when an ac magnetic field is applied. Consider the process of applying an ac magnetic field $\mathbf{B}=B_x\hat{\mathbf{x}}$ as shown in Fig.~\ref{fig2}a. A circulating electric field $\mathbf{E}$ parallel to side surface (parallel to $\mathbf{B}$) is generated due to Faraday's law of induction, where $\mathbf{E}^t=-\mathbf{E}^b=(\partial B_x/\partial t)(d/2)\hat{\mathbf{y}}$. Such an electric field will induce a Hall current density $\mathbf{j}^{\text{2D}}=\mathbf{j}_t^{\text{2D}}+\mathbf{j}_b^{\text{2D}}$, where $\mathbf{j}_t^{\text{2D}}=\sigma^t_{xy}\hat{\mathbf{z}}\times\mathbf{E}^t$ and $\mathbf{j}_b^{\text{2D}}=\sigma^b_{xy}\hat{\mathbf{z}}\times\mathbf{E}^b$. Therefore the total current $\boldsymbol{\mathcal{J}}=\mathbf{j}^{\text{2D}}\ell=\mathcal{J}\hat{\mathbf{x}}$, where
\begin{equation}\label{current}
\mathcal{J}=\frac{\theta}{\pi}\frac{e^2}{2h}\frac{\partial B_x}{\partial t}\ell d.
\end{equation}
Here $d$ and $\ell$ are the thickness and width of the TI film as shown in Fig.~\ref{fig2}a, and $\theta\rightarrow\pi$ when $d$ is large enough. For an estimation, take $B_x=B_0e^{-i\omega t}$, $B_0=10$~G, $\omega/2\pi=1$~GHz, $d=20$~nm, $\theta/\pi\approx0.91$ (finite-size effect taken into account as in Fig.~\ref{fig3}), and $\ell=500$~$\mu$m, we have $\mathcal{J}=-i\mathcal{J}_0e^{-i\omega t}$ with $\mathcal{J}_0=1.11$~nA, in the range accessible by transport experiments. Moreover, as shown in Fig.~\ref{fig3}b, the current amplitude $\mathcal{J}_0$ scales linearly with thickness $d$, for $\theta$ is a linear function of $1/d$ with $(1-\theta/\pi)\propto 1/d$, i.e., thicker film gives rise to larger TME.

\paragraph{Finite-size effect}

Due to the finite-size confinement along $z$ direction, the TME effect is not quantized when the TI film is thin. However, as we shall show below, the TME coefficient converges quickly into the quantized value as the film thickness $d$ increases. The generic Hamiltonian of a TI thin film can be written as $\mathcal{H}_{\text{2D}}(\mathbf{k})=\int_{-d/2}^{d/2}dz\mathcal{H}_{\text{3D}}(\mathbf{k},z)$. Here $\mathbf{k}=(k_x, k_y)$, and we impose periodic boundary conditions in both $x$ and $y$ directions. The magnetoelectric response of such a thin film can be directly calculated with the Kubo formula. With the 3D in-plane current density operator defined as $\mathbf{j}^{\text{3D}}(\mathbf{k},z)=(e/\hbar)\partial_\mathbf{k}\mathcal{H}_{\text{3D}}(\mathbf{k},z)$, we can write down a dc current correlation function
\begin{eqnarray}
\Pi_{xy}&&(z,z')=\frac{\hbar^2}{2\pi e^2}\int d^2\mathbf{k}\sum_{n\neq m}f(\epsilon_{n\mathbf{k}})
\nonumber
\\
&&\times2\text{Im}\left[\frac{\left\langle u_{n\mathbf{k}}\right|j^{\text{3D}}_x(\mathbf{k},z)\left|u_{m\mathbf{k}}\right\rangle\left\langle u_{m\mathbf{k}}\right|j^{\text{3D}}_y(\mathbf{k},z')\left|u_{n\mathbf{k}}\right\rangle}{(\epsilon_{n\mathbf{k}}-\epsilon_{m\mathbf{k}})^2}\right],
\nonumber
\end{eqnarray}
where $|u_{n\mathbf{k}}\rangle$ is the normalized Bloch wavefunction in the $n$-th electron subband satisfying $\mathcal{H}_{\text{2D}}(\mathbf{k})\left|u_{n\mathbf{k}}\right\rangle=\epsilon_{n\mathbf{k}}|u_{n\mathbf{k}}\rangle$ , and $f(\epsilon)$ is the Fermi-Dirac distribution function. The Kubo formula for magnetic field $B_y$ induced by a uniform external electric field $E_y$ is then given by
\begin{equation}
B_y(z)=-\mu \frac{e^2}{2h}\kappa(z) E_y\ ,
\end{equation}
where $\kappa(z)=\int_{-d/2}^{d/2}dz_1\ \mathrm{sgn}(z-z_1)\int_{-d/2}^{d/2}dz_2\ \Pi_{xy}(z_1,z_2)$ is a dimensionless function. Here $\mathrm{sgn}(z)$ gives the sign of $z$. Similarly, the current density $j^{\mathrm{3D}}_x$ induced by a uniform external ac magnetic field $B_x$ of frequency $\omega/2\pi$ is given by
\begin{equation}\label{3Dcurrent}
j^{\mathrm{3D}}_x(z)=-i\omega\frac{e^2}{2h}\eta(z) B_x\ ,
\end{equation}
where $\eta(z)=2\int_{-d/2}^{d/2}dz_1\ z_1\Pi_{xy}(z,z_1)$ is also dimensionless.

The response formulas above are generic for any TI system and do not rely on a specific model. For concreteness, we adopt the effective Hamiltonian in Ref.~\cite{wang2013a} to describe the low-energy bands of Bi$_2$Te$_3$ family materials,
$\mathcal{H}_{\mathrm{3D}}(\mathbf{k},z)=\varepsilon 1\otimes1+d^1\tau_1\otimes1+d^2\tau_2\otimes\sigma_3+d^3\tau_3\otimes1-\Delta(z)\tau_3\otimes\sigma_3+iA_1\partial_z\tau_2\otimes\sigma_2$. Here $\tau_j$ and $\sigma_j$ ($j=1,2,3$) are Pauli matrices, $\varepsilon(\mathbf{k},z)=-D_1\partial_z^2+D_2(k_x^2+k_y^2)$, $d^{1,2,3}(\mathbf{k},z)=(A_2k_x, A_2k_y, B_0-B_1\partial_z^2+B_2(k_x^2+k_y^2))$, and $\Delta(z)$ is the $z$-dependent exchange field. We then discretize it into a tight-binding model along $z$-axis between neighboring quintuple layers (QL) from $\mathcal{H}_{\text{3D}}$, and assume $\Delta(z)$ takes the values $\pm\Delta_s$ in the top and bottom layers, respectively, and zero elsewhere. Fig.~\ref{fig2}c shows the numerical calculations of $\kappa(z)$ and $\eta(z)$ for thin films of $6$, $10$ and $20$~QL, where we set a typical surface exchange field $\Delta_s=50$~meV. All the other parameters are taken from Ref.~\cite{zhang2009} for (Bi$_{0.1}$Sb$_{0.9}$)$_2$Te$_3$. The bulk value of $\kappa(z)$ at $z=0$ as a function of $d$ is plotted as black line in Fig.~\ref{fig3}. As is consistent with the topological field theory, $\kappa(z)$ in the bulk tends to $1$ and becomes quantized as the thickness $d$ increases, whereas $\eta(z)$ is bounded within a finite penetration depth to the top and bottom surfaces. The shape of functions $\kappa(z)$ and $\eta(z)$ near surfaces remain almost unchanged as the thickness $d$ varies.

To characterize the deviation from topological quantization of TME in TI thin films, we further define the dimensionless number
\begin{equation}
\gamma=\frac{1}{d}\int_{-d/2}^{d/2} dz\ \kappa(z)=\frac{1}{d}\int_{-d/2}^{d/2} dz\ \eta(z)\ ,
\end{equation}
which is the mean value of $\kappa(z)$ or $\eta(z)$ (which are equal to each other). The average magnetic field in response to the external electric field $E_y$ is then $B^{\text{mean}}_y=-\gamma\mu(e^2/2h)E_y$, whereas the total 2D current density induced by external magnetic field $B_x$ is given by $j_x^{2D}=\int dz j^{3D}_x(z)=-i\omega\gamma d(e^2/2h)B_x$. Compared to Eq.~(\ref{current}), we get $\theta/\pi=\gamma$. The value of $\gamma$ as a function of $d$ is shown as the red line in Fig.~\ref{fig3}, where $\gamma\rightarrow1$ with $d\rightarrow\infty$. This shows the TME effect is quantized as the system is in the thermodynamic limit. In fact, as shown in the inset of Fig.~\ref{fig3}, the value of $1-\gamma$ scales linearly with $1/d$ as the thickness $d\rightarrow\infty$, which indicates $\int_{-d/2}^{d/2} dz(1-\kappa(z))=\text{const.}$ when $d$ is large enough. This is simply because the function $1-\kappa(z)$ is nonzero only near the top and bottom surfaces, and its shape is independent of the thickness $d$, as shown in Fig.~\ref{fig2}c. Since $(1-\gamma)\propto 1/d$, $\mathcal{J}_0\propto\theta d=\gamma\pi d$ is a linear function of $d$, as is shown in Fig.~\ref{fig3}b.

\paragraph{Discussion}

\begin{figure}[t]
\begin{center}
\includegraphics[width=3.2in]{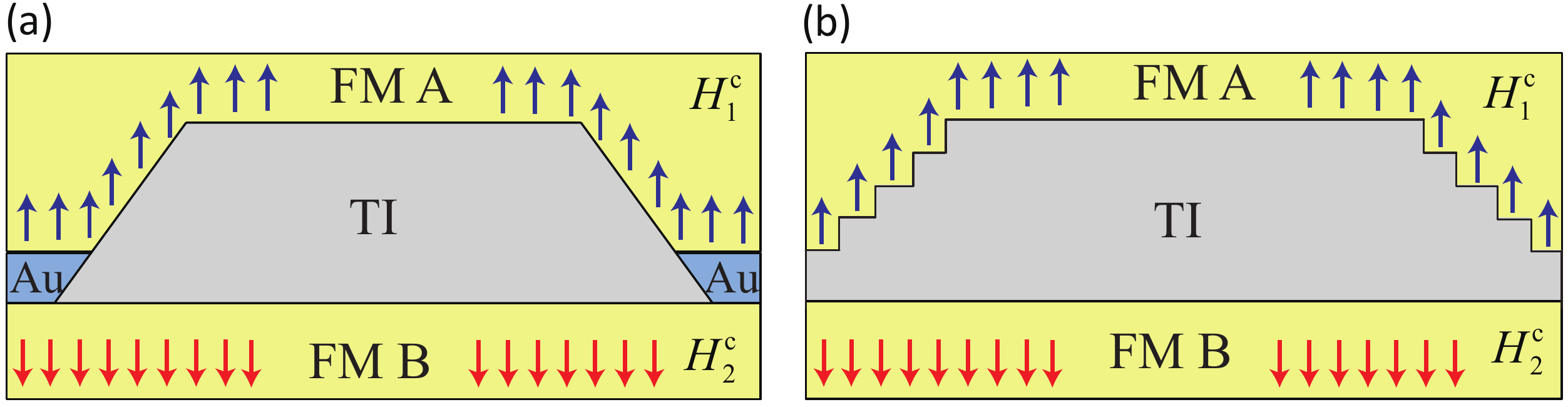}
\end{center}
\caption{(color online). Schematic of FM-TI-FM heterostructure to observe the quantized TME, where the side surface of TI is gapped by FM proximity in (a). Au is the electrode. Such geometry of TI can be made by lithography. In reality, such configuration cook in experiments may like (b), where the side surface is uneven and step-like. In this case, the side surface is gapped by both FM ordering and quantum confinement.}
\label{fig4}
\end{figure}

The TME effect in the setup Fig.~\ref{fig2} is not quantized when the side surface (parallel to $\hat{\mathbf{z}}$) is not gapped. The gapless side surface states may give rise to a non-topological contributions to the TME effect. To eliminate such non-topological contribution, one can use the device setup as shown in Fig.~\ref{fig4}, where the side surface is gapped either by FM order or quantum confinement. Also in this setup, if FM A and B have opposite sign of exchange coupling parameter to TI surface, only parallel magnetization is needed to realize TME. Recently, the surface QHE has been realized in bulk Bi$_{2-x}$Sb$_x$Te$_{3-y}$Se$_y$ TI~\cite{xu2014,yoshimi2015}, where the systems exhibit surface-dominated conduction even at temperatures close to the room temperature, whereas the bulk conduction is negligible. Such experimental progress on the material growth and rich material choice of TI and FM insulator make the realization of the quantized TME in TIs feasible.

\begin{acknowledgments}
We are grateful to Ke He for sharing their data prior to publication, and thank Yihua Wang and Andre Broido
for useful comments on the draft. This work is supported by the US Department of Energy, Office of Basic Energy Sciences, Division of Materials Sciences and Engineering, under Contract No.~DE-AC02-76SF00515 and in part by the NSF under grant No.~DMR-1305677. XLQ is supported by the NSF through the grant No.~DMR-1151786.
\end{acknowledgments}

{\it Note added}: During the preparation of our manuscript, we learned of an independent work on a similar problem~\cite{morimoto2015}. However, their experimental proposal to observe the TME is different from our results.


\begin{thebibliography}{30}
\bibitem [{uni()}]{unit}%
  \BibitemOpen
  \href@noop {} {}\bibinfo {note} {We use SI units. For CGS units
  $e^2/h\rightarrow \alpha/2\pi$, where $\alpha$ is the fine structure
  constant.}\BibitemShut {Stop}%
\end{thebibliography}
\end{document}